\documentclass[3p,times]{elsarticle}

\usepackage{ecrc}

\volume{00}

\firstpage{1}

\journalname{Nuclear Physics A}

\runauth{}

\jid{npa}

\jnltitlelogo{Nuclear Physics A}

\usepackage{amssymb}

\biboptions{square,comma,numbers,sort&compress}

\usepackage[figuresright]{rotating}

\begin{document}

\begin{frontmatter}



\dochead{}

\title{Dynamics of strongly interacting parton-hadron matter}


\author{W. Cassing}
\author{O. Linnyk}

\address{Justus Liebig University of Giessen,
  35392 Giessen,
  Germany }

\author{E. L. Bratkovskaya}

\address{ Johann Wolfgang Goethe University,
 60438 Frankfurt am Main,
 Germany}

\begin{abstract}
In this study we investigate the dynamics of strongly interacting
parton-hadron matter by calculating the centrality dependence of
direct photons produced in Au+Au collisions at
$\sqrt{s_{NN}}=200$~GeV
within the Parton-Hadron-String Dynamics (PHSD) transport approach.
As sources for 'direct' photons, we incorporate the interactions of
quarks and gluons as well as hadronic interactions
($\pi+\pi\to\rho+\gamma$, $\rho+\pi\to\pi+\gamma$, meson-meson
bremsstrahlung $m+m\to m+m+\gamma$, meson-baryon bremsstrahlung
$m+B\to m+B+\gamma$), the decays of $\phi$ and $a_1$ mesons and the
photons produced in the initial hard collisions ('pQCD'). Our
calculations suggest that the channel decomposition of the observed
spectrum changes with centrality with an increasing (dominant)
contribution of hadronic sources for more peripheral reactions.
Furthermore, the 'thermal' photon yield is found to scale roughly
with the number of participant nucleons as $N_{part}^\alpha$ with
$\alpha \approx$ 1.5, whereas the partonic contribution scales with
an exponent  $\alpha_p \approx1.75$. Additionally, we provide
predictions for the centrality dependence of the direct photon
elliptic flow $v_2(p_T)$. The direct photon $v_2$ is seen to be
larger in peripheral collisions compared to the most central ones
since the photons from the hot deconfined matter in the early stages
of the collision carry a much smaller elliptic flow than those from the final
hadronic interactions.
\end{abstract}


\end{frontmatter}

\section{Introduction}
\label{I1} The 'direct photons' from relativistic heavy-ion
collisions are expected to be a valuable probe of the collision
dynamics at early times and to provide information on the
characteristics of the initially created parton-hadron matter once
the final state hadronic decay photons are subtracted from the
experimental spectra~\cite{Shuryak:1977ut,Peitzmann:2001mz}. In the
last years, the PHENIX Collaboration~\cite{PHENIX1,PHENIXlast} has
measured the spectra of the photons produced in minimal bias Au+Au
collisions at $\sqrt{s_{NN}}=200$~GeV and found a strong elliptic
flow $v_2(p_T)$ of 'direct photons', which is comparable to that of
the produced pions. Since direct photons were expected to be
essentially produced in the initial hot medium before the collective
flow has developed, this observation was in contrast to the
theoretical expectations and predictions \cite{Chatterjee:2005de}.
Also more recent studies employing  event-by-event hydrodynamical
calculations~\cite{Chatterjee:2013naa,Shen:2013cca} severely have
underestimated the elliptic flow of direct photons and alternative
sources of direct photons from the conformal anomaly have been
suggested~\cite{Basar:2012bp,Bzdak:2012fr}.

On the other hand, in Refs.~\cite{Linnyk:2013hta,Linnyk14} we have
proposed that apart from the partonic production channels the direct
photon yield and primarily the strong $v_2$ might be due to hadronic
sources (such as meson-meson Bremsstrahlung or hadronic interactions
as $\pi+\pi\to\rho+\gamma$, $\rho+\pi\to\pi+\gamma$ etc.).  Indeed,
the interacting hadrons carry a large $v_2$ and contribute by more
than 50\% to the measured 'direct photons' in minimum bias
collisions at RHIC according to the PHSD calculations in
Ref.~\cite{Linnyk:2013hta} (cf. also the hydrodynamics calculations
in Ref.~\cite{Dusling:2009ej}). For a quantitative understanding of
the direct photon production it is important to verify the
decomposition of the total photon yield according to the production
sources: the late hadron decays (the cocktail), hadronic
interactions beyond the cocktail (during the collision phase) and
the partonic interactions in the quark-gluon plasma (QGP). Since
previous transport studies have indicated that the duration of the
partonic phase substantially decreases with increasing impact
parameter \cite{PHSDasymmetries} we will study here explicitly the
centrality dependence of the direct photon yield together with the
essential production channels and their impact on the photon $v_2$.

As in Ref.~\cite{Linnyk:2013hta} we will employ the
Parton-Hadron-String Dynamics (PHSD) transport approach to
investigate the photon production in Au+Au collisions at
$\sqrt{s_{NN}}=200$~GeV at various centralities thus extending the
previous investigations for the case of minimum bias collisions
(see also \cite{Linnyk14}. We recall that the PHSD approach has provided a
consistent description of the bulk properties of heavy-ion
collisions -- rapidity spectra, transverse mass distributions,
azimuthal asymmetries of various particle species -- from low
Super-Proton-Synchrotron (SPS) to top
Relativistic-Heavy-Ion-Collider (RHIC) energies \cite{PHSDasymmetries} and was successfully
used also for the analysis of dilepton production from hadronic and
partonic sources at SPS, RHIC and Large-Hadron-Collider (LHC)
energies ~\cite{Linnyk2011_BOTH}. It is therefore of interest to
calculate also the photon production in relativistic heavy-ion
collisions from  hadronic and partonic interactions within the PHSD
transport approach, since its microscopic and non-equilibrium
evolution of the nucleus-nucleus collision is independently
controlled by a multitude of other hadronic and electromagnetic
observables in a wide energy range~\cite{BrCa11,PHSDasymmetries,Linnyk2011_BOTH}.

\section{Photons within PHSD}
For the details on the PHSD approach we refer the reader to Refs.
\cite{BrCa11,CasBrat} and the implementation of the photon
production to Refs.~\cite{ElenaKiselev,Linnyk:2013hta} (and
references therein). Let us recall that the dynamical calculations
within the PHSD for dileptons  agree with the dilepton rate emitted by the thermalized
QCD medium as calculated in  lattice QCD (lQCD)~\cite{Linnyk2011_BOTH}. We note,
additionally, that the electric conductivity of the QGP from the
PHSD, which controls the photon emission rate in equilibrium, is
rather well in line with available lQCD results \cite{Cassing:2013iz}.

As sources of photon production - on top of the general dynamical
evolution - we consider hadronic~\cite{ElenaKiselev,Gale87} as well
as partonic~\cite{Feinberg:1976ua} interactions. In the present
study we extend the calculations in Ref.~\cite{Linnyk:2013hta} by
adding an additional source of photons, i.e. the bremsstrahlung in
elastic meson+baryon collisions ($m+B\to m+B+\gamma$). This process
is calculated within the soft photon approximation for charged
hadrons in analogy to the treatment of photon production by the
meson+meson bremsstrahlung in
Refs.~\cite{Gale87,ElenaKiselev,Linnyk:2013hta,Linnyk14}.

\begin{figure}
\resizebox{1.0\textwidth}{!}{%
 \includegraphics{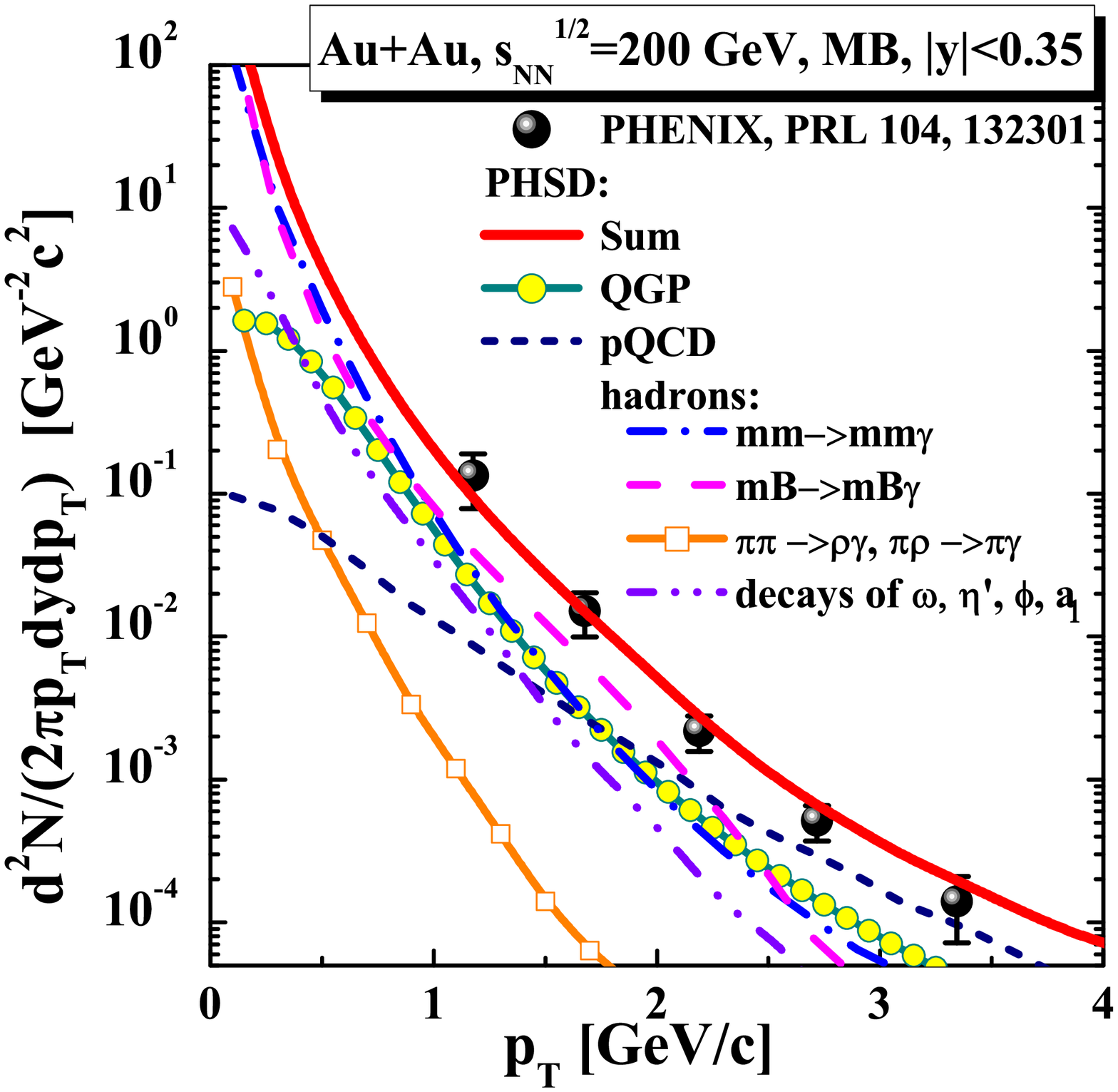}  \includegraphics{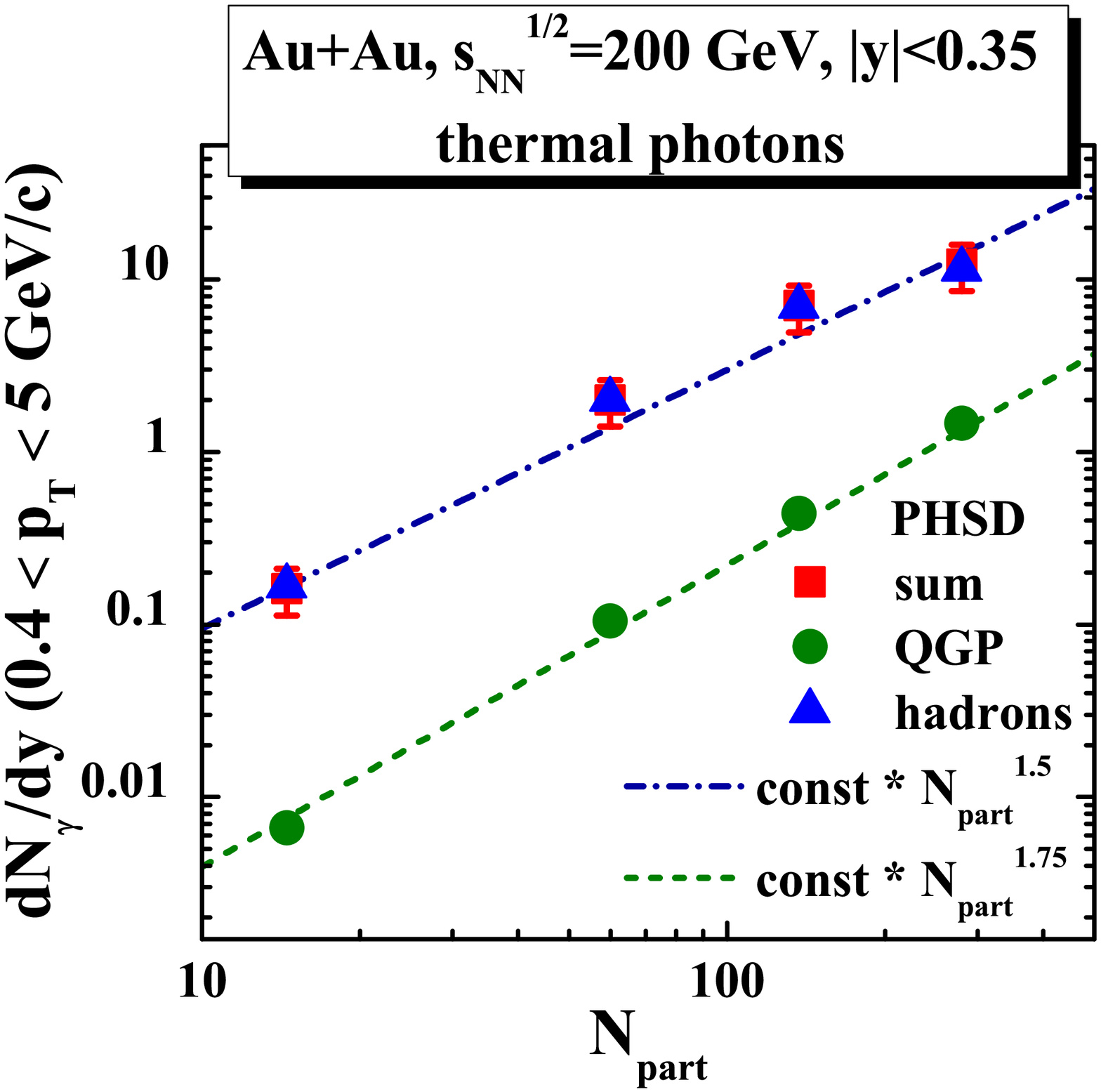}
} \caption{(lhs) Direct photons (sum of all photon production
channels except the $\pi$- and $\eta$-meson decays) from the PHSD
approach (red solid line) in comparison to the data of the PHENIX
Collaboration~\protect{\cite{PHENIXlast} for minimal bias collisions
of Au+Au at $\sqrt{s_{NN}}=200$~GeV } (black symbols). The various
channels are described in the legend. (rhs) Integrated spectra of
thermal photons (full squares) produced in Au+Au collisions at
$\sqrt{s_{NN}}=200$~GeV versus
 the number of participants
$N_{part}$. The scaling with $N_{part}$ from the QGP contribution
(full dots) and the bremsstrahlungs channels (full triangles) are
shown separately.} \label{MB}
\end{figure}

Since a new production mechanism has been added to the hadronic
production channels, we first check whether this addition does not
lead to an overestimation of the data from the PHENIX
Collaboration~\cite{PHENIXlast}  in minimal bias $Au+Au$ collisions
in Fig.~\ref{MB} (l.h.s.). Since the decays of mesons as 'late'
hadronic sources are less sensitive to the creation of the hot and
dense medium and to its properties, they are usually subtracted
experimentally from the total photon yield to access the `direct'
photon spectrum. In our calculations of the direct photon spectrum
in Fig.~\ref{MB} (l.h.s.) the following sources are taken into account: the
decays of $\omega$, $\eta$', $\phi$ and $a_1$ mesons; the reactions
$\pi+\rho\to\pi+\gamma$, $\pi+\pi\to \rho+\gamma$; the photon
bremsstrahlung in meson-meson and meson-baryon collisions $m+m\to
m+m+\gamma$, $m+B\to m+B+\gamma$; photon production in the QGP in
the processes $q+{\bar q}  \to g+\gamma$, and $q({\bar q})+g \to
q({\bar q})+\gamma$ as well as the photon production in the initial
hard collisions ("pQCD"), which is given by the hard photon yield in
p+p collisions scaled with the number of binary collisions
$N_{coll}$. We find that our PHSD calculations are in a reasonable
agreement with the PHENIX data~\cite{PHENIXlast} and show a clear
dominance of the hadronic production channels over the partonic
channels for transverse momenta below about 0.7 GeV/c. In
particular, the bremsstrahlung contributions are responsible for the
'banana shape' spectrum and the strong increase for low $p_T$.
Accordingly, especially experimental data well below 1 GeV/c in
$p_T$ will be helpful in disentangling the various sources.

We have, furthermore, calculated the produced photons as a function
of the number of participants $N_{part}$. Integrating the thermal
photon spectra - defined by the partonic radiation channels and the
hadronic non-decay contributions -  over the transverse momentum
$p_T$ in the interval $0.4 \leq p_T \leq 5$GeV/c, we obtain the
number of 'thermal photons'  as a function of centrality, which is
plotted in Fig.~1 (r.h.s.) (full squares) as a function of
$N_{part}$ calculated in the Monte-Carlo Glauber model described in
Ref.~\cite{phenMCGlau}. Since only binary collision channels
contribute to the production of thermal photons in our approach,
their yield rises faster than $N_{part}$. A power-law fit to our
results gives approximately a scaling $\sim N_{part}^\alpha$ with
$\alpha \approx$ 1.5. In addition we display in Fig.~1 (r.h.s.) the
scaling with $N_{part}$ for the partonic (full dots) and hadronic
bremsstrahlung channels (full triangles) separately, which give
exponents of $\approx$ 1.75 and $\approx$ 1.5, respectively.

\begin{figure}
\resizebox{1.0\textwidth}{!}{%
 \includegraphics{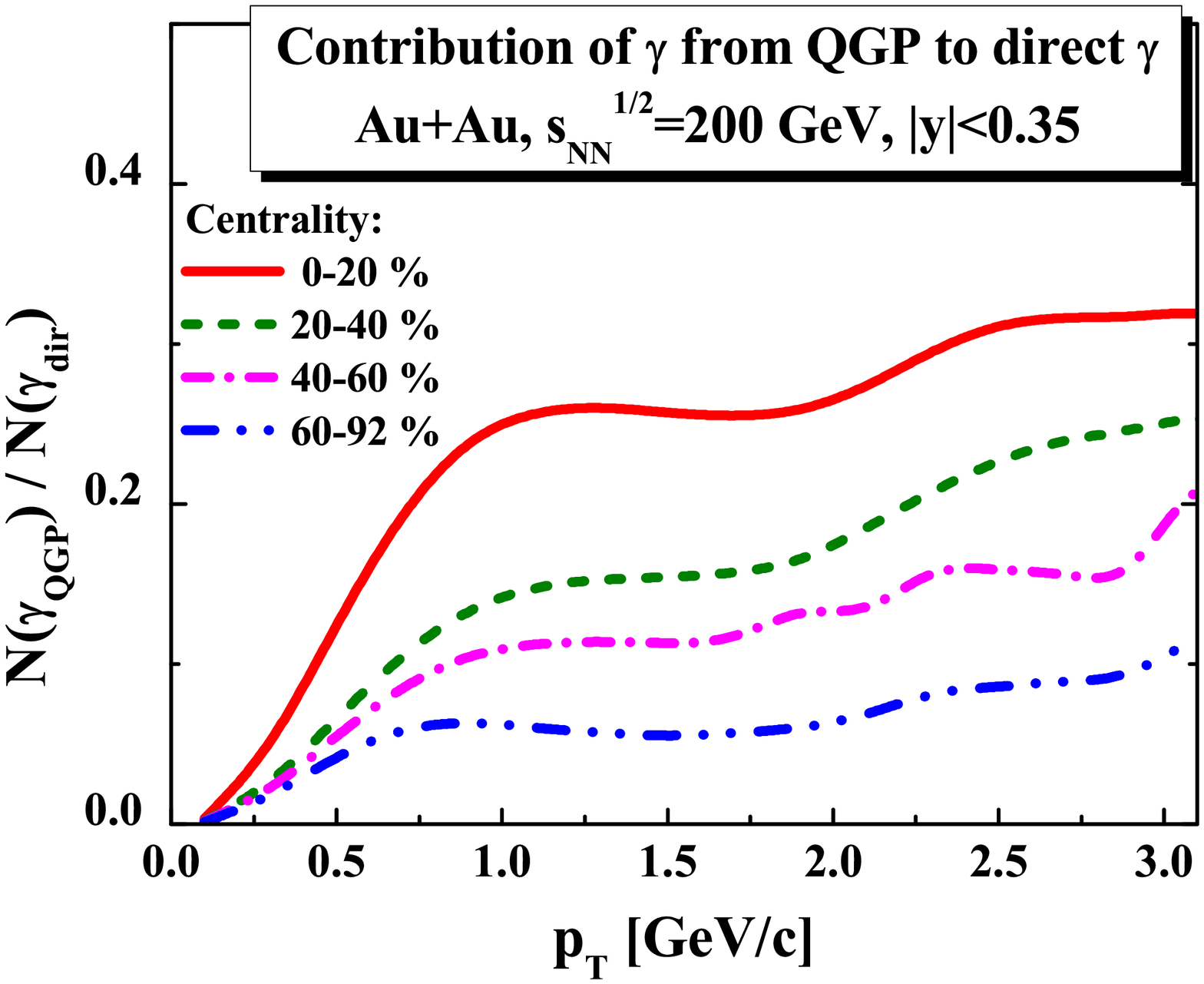}  \includegraphics{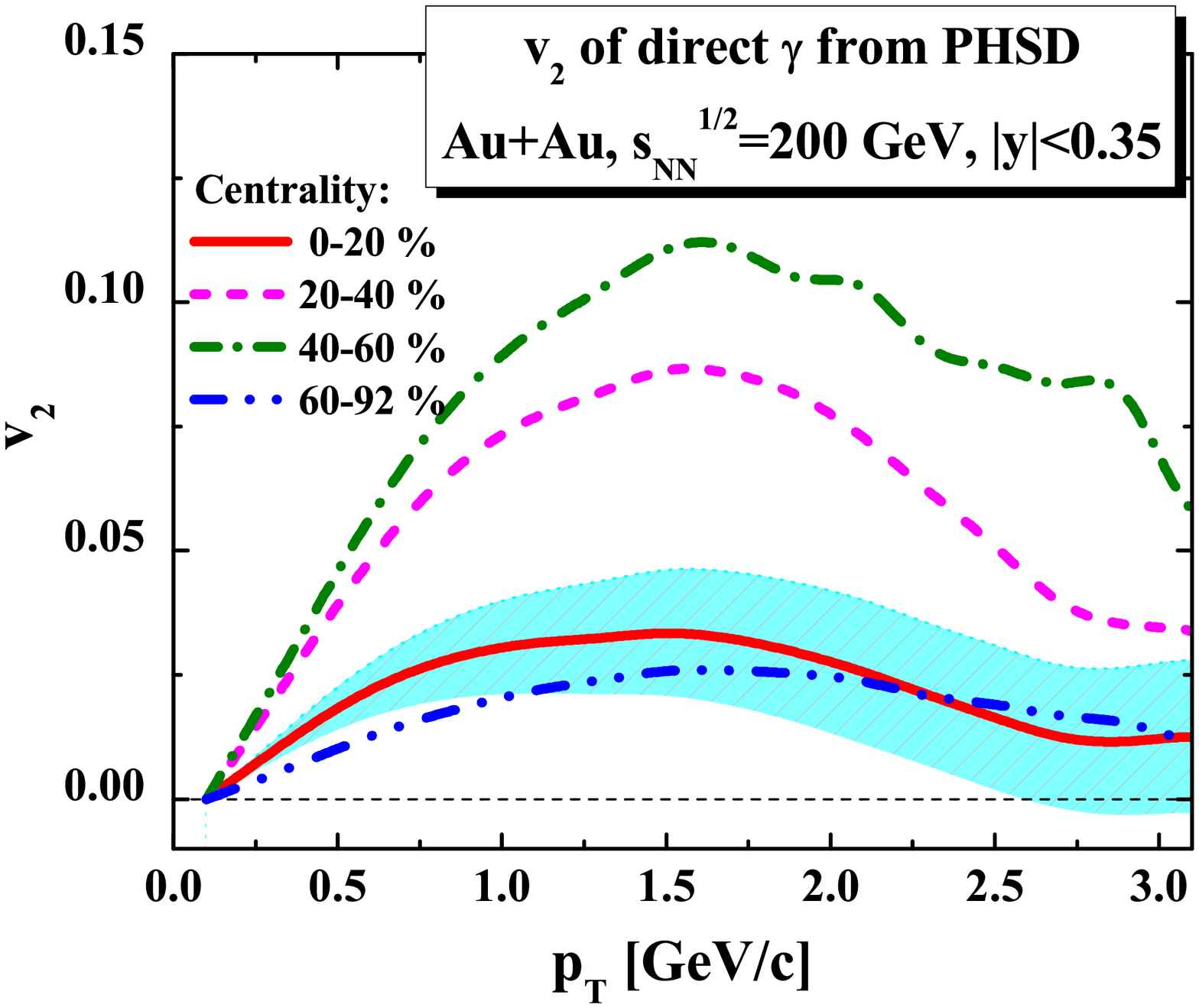}
} \caption{(lhs) The ratios of the number of photons produced in the
QGP to the number of all direct photons produced through binary
processes in different-centrality Au+Au collisions at
$\sqrt{s_{NN}}=200$~GeV versus the photon transverse momentum $p_T$.
(rhs) The elliptic flow $v_2(p_T)$ of direct photons produced
through binary processes in  Au+Au collisions at
$\sqrt{s_{NN}}=200$~GeV for different centralities versus the photon
transverse momentum $p_T$. The hatched area (for the most central
bin) stands for the statistical uncertainty in the photon $v_2$ from
PHSD which in width is also characteristic for the other
centralities.} \label{ratio}
\end{figure}

As one might have expected  the contribution of the photons from the
QGP is larger in central collisions in PHSD while the hadronic sources
contribute more dominantly in peripheral collisions. We quantify the
relative contributions by plotting in Fig.~\ref{ratio} (l.h.s.) the
ratio of the number of photons produced in the QGP to the number of
all direct photons (from the QGP, $m+m/B\to m+m/B+\gamma$,
$\pi+\pi/\rho\to\rho/\pi+\gamma$ and the pQCD photons). The
contribution of the QGP photons is seen to increase with transverse
momentum and reaches slightly more than 30\% for the most central
event bin. On the other hand, the ratio of QGP photons to the total
direct photons falls rapidly with decreasing centrality and is below
10\% in the most peripheral centrality bin. Accordingly, minimal
bias collisions are dominated by the hadronic channels that come
along with a large hadronic elliptic flow $v_2$.

In Fig.~\ref{ratio} (r.h.s.)  we provide predictions for the
centrality dependence of the direct photon elliptic flow $v_2(p_T)$
within the PHSD approach. The direct photon $v_2$ is seen to be
larger in the peripheral collisions compared to the most central
ones. This result can be readily understood when keeping in mind the
ratios presented in Fig.~\ref{ratio} (l.h.s.) and the roughly linear increase
of the hadron $v_2$ with impact parameter \cite{PHSDasymmetries}. As
has been described in detail in Ref.~\cite{Linnyk:2013hta}, the PHSD
approach predicts a very small $v_2$ of photons produced in the
initial hot deconfined phase by partonic channels  of the order of
2\%.

\section{Summary}

The spectra of direct and thermal photons - as produced in Au-Au
collisions at $\sqrt{s_{NN}}=200$~GeV - have been calculated
differentially in collision centrality within the PHSD transport
approach, which has been previously  tested in comparison to the
measured spectra and flow of photons in minimal bias collisions at
the same energy~\cite{Linnyk:2013hta}. We have found that the
channel decomposition of the photon spectra changes with centrality
providing a larger contribution of the hadronic sources in more
peripheral collisions. As a consequence, the direct photon $v_2$ is
larger in peripheral collisions as compared to the most central
reactions. This is due to the much smaller elliptic flow of the
photons from the hot deconfined matter in the early stages of the
collision relative to the $v_2$ from final hadrons within our
approach. The increase of the direct photon $v_2$ with decreasing
centrality for the two most central bins has been also indicated in
hydrodynamics calculations in Refs.~\cite{Shen:2013cca}, although
with slightly lower absolute values of $v_2$. Future measurements of
the photon spectra and elliptic flow as a function of the collision
centrality will be mandatory for a clarification of the 'photon
$v_2$ puzzle' from the experimental side and to estimate the
contribution from unconventional sources
\cite{Bzdak:2012fr,Basar:2012bp}.

Furthermore, since only collisional channels contribute to the
production of thermal photons in PHSD, their yield rises faster than
the number of participating nucleons $N_{part}$ as expected also
from qualitative considerations. A power-law fit to our results
gives approximately a scaling $\sim N_{part}^\alpha$ with $\alpha
\approx$1.5, whereas the partonic and hadronic channels separately
scale with exponents of $\approx$ 1.75 and $\approx$ 1.5,
respectively.

\end{document}